

\pagewidth{6.2in}
\pageheight{9.5in}
\baselineskip=20pt

\heading A TOPOLOGICALLY STABLE SOLUTION \\ IN QUANTUM ELECTRODYNAMICS.
\endheading
\heading by Edwin J. Beggs,\\University College, Swansea,\\Wales SA2 8PP.
\endheading
\heading Abstract
\endheading

\newpage

\heading A TOPOLOGICALLY STABLE SOLUTION \\ IN QUANTUM ELECTRODYNAMICS.
\endheading
\heading by Edwin J. Beggs,\\University College, Swansea,\\Wales SA2 8PP.
\endheading
\heading Introduction
\endheading

It has been remarked to
me that it is curious that it is sufficient to use trivial bundles in QED.
The purpose of this paper is to construct some exact solutions to the equations
of quantum electrodynamics, treated as a classical field theory. These
solutions exhibit topological stability, that is
their existence is linked with a non-trivial bundle structure on space-time,
and so they cannot be continuously
deformed to the zero solution.  In this respect they are similar to instantons
[1,2]. The eventual aim would be to
represent particles as solitons in the theory.

 Electrons behave in a very strange way under rotation. One
complete revolution, and they change sign. In other words, they are Fermions, a
concept well understood in quantum theory, but not classically. Let us suppose
that the spinor field $\psi(x)$ is well
defined and single valued on space-time. If we take $\psi(x)$ and rotate $x$ by
one revolution, we get $-\psi(x)$, and
deduce that $\psi(x)=0$. There is a way out of this, and that is to assume that
$\psi$ is
not a function of $x$, but rather lives on a non-trivial bundle over $3+1$
dimensional space
time. This form of Kaluza-Klein theory is the approach
that we shall take in trying to construct the electron function. We shall find
that going to a higher dimensional model
also allows the existence of solutions with charges of both signs. This is
important as the Dirac sea argument only
works for quantum particles, not classical ones. In a classical theory, if
there is an electron, there should also be an
explicit positron. In the quantum field theory in 3+1 dimensions, this problem
is neatly side-stepped by the operator
valued nature of everything effectively allowing negative charge density.  The
fibers in the theory have signature
$1+1$, and are non-compact. For other theories with non-compact fiber spaces,
the reader can consult [10, 11].

The solutions have topological singularities at their centers. The conserved
currents vanish everywhere except at the
singular points. That is, the charge of the solutions arises entirely
topologically, and so is automatically quantised by
the type of singularity. The electromagnetic field is that of a point electric
charge. The quantisation of the charge is
similar to that for magnetic monopoles [6].

There is no claim that this is the `correct' classical electron solution, in
fact there are many places where the
construction could have been altered. However a solution of very simple form
does exist, and is topologically stable,
as is the oppositely charged solution. Since its charge is inversely
proportional to the coupling constant $q$, it is likely
that the solution would be difficult to access via perturbation theory. The
electromagnetic field is well defined on the
usual $3+1$ dimensional space time, but the spinor field varies by a complex
multiple over the fiber space. The fact that
the theory is 4+2 dimensional may be controversial, but there are many other
modern theories which rely on
compactified extra dimensions. The reader should also note that Dirac [4]
showed that conformal transformations in
3+1 dimensions correspond to isometries in 4+2 dimensions. It is the author's
opinion that this is not entirely unrelated
to the existence of the solutions described here.

It is hoped that these solutions
may be of interest in indicating what other constructions may be possible in
classical field theory, or at least have
some curiosity value!

I should like to thank D. Olive, P. Johnson, N. Landsman, S. Majid and D.A.
Dubin for several conversations.

\heading Contents
\endheading

\underbar {Section 1}\quad The field equations of Quantum Electrodynamics.

\underbar {Section 2}\quad The Dirac equation for the Electron.

\underbar {Section 3}\quad The conserved currents.

\underbar {Section 4}\quad The quantum numbers.

\heading Section 1 \\The field equations of Quantum Electrodynamics.
\endheading

We take the Lorentz group to act on vectors by the formula
$x^a \mapsto \Lambda^a_{\phantom{a}b} x^b$.
The condition that the metric $g_{ab}$ is preserved is
that $g_{ab} \Lambda^a_{\phantom{a}c} \Lambda^b_{\phantom{b}d}=g_{cd}$.
If we move to the Lie algebra, we
write $ \Lambda^a_{\phantom{a}b} = g^a_{\phantom{a}b} +
\omega^a_{\phantom{a}b}$,  and the metric preserving condition
is then that $\omega_{ab}=-\omega_{ba}$. The Lorentz group has fundamental
group
$\Bbb Z_2$, and the corresponding double cover is isomorphic to $SL_2(\Bbb C)$.
 The symbol $\partial
_\mu$ is the differential operator $\frac {\partial}{\partial x^\mu}$.

Electrodynamics is an abelian gauge theory, and we have the gauge group $S^1$.
Thus the connection can be regarded as an imaginary number valued
1-form, which we denote $iA_\mu$. The corresponding covariant derivative is
$D_\mu = \partial_\mu - iqA_\mu$, where $q$ is a real coupling constant. The
curvature is defined to be $F_{\mu\nu}=\partial_\mu A_\nu - \partial_\nu
A_\mu$.

The massless Dirac equation for the spinor field $\psi$ can now be written as
$$\gamma^\mu D_\mu \psi\ =\ \gamma^\mu \big(\partial_\mu-iqA_\mu\big) \psi\ =\
0\ .\tag 1.1$$
The double cover of the Lorentz group acts on the Dirac spinors in the
following way. (We only give the Lie algebra representation).
$$(\omega)(\psi)\ =\ \tfrac 18  \omega_{ab}.\big[ \gamma^a ,\gamma^b
\big].\psi\
.$$

Maxwell's equations can be written in terms of the current $j_\nu=q \bar\psi
\gamma_\nu \psi$ as
$$\partial^\mu F_{\mu\nu}\ =\  j_\nu\ = \ q \bar\psi
\gamma_\nu \psi\ .\tag 1.2$$
In 6 dimensional space-time with signature $(+----+)$ (for indices 0 to 5) the
formula $\bar\psi=\psi^*\gamma^0$
(where $*$ denotes
conjugate transpose)
simply does not give a conserved current (as can easily be seen by direct
calculation from the Dirac equation).
However, the definition
$\bar\psi=\psi^*\gamma^1\gamma^2\gamma^3\gamma^4$ does give a conserved
current. The matrix
$\gamma^1\gamma^2\gamma^3\gamma^4$ has the property that it commutes with
all the Hermitian gamma matrices ($\gamma^0$ and $\gamma^5$), and anticommutes
with all the
anti-Hermitian gamma matrices ($\gamma^1$, $\gamma^2$, $\gamma^3$ and
$\gamma^4$).

If we were in a space-time with only one timelike coordinate, the matrix with
these commutation properties would
be $\gamma^0$.
Then the timelike current (charge density) would
be positive definite, which would prevent the similtaneous existence of
solutions with charges of both signs in a
classical theory. For other applications of a theory with 2 timelike
directions, see [7].

\heading Section 2 \\The Dirac equation for the Electron.
\endheading

We wish to construct an electron at rest on the time ($x^0$) axis. To do this
we try to get a solution
which is as spherically symmetric as possible. If $\psi$ is a spherically
symmetric spinor field, then we should like to satisfy the equation
$$\psi(g(p))\ =\ g.\psi(p)\ ,$$
where $p$ is a position in space time, for some $g\in SL_2$. Here $g$ should
preserve any obvious invariants associated with the electron. The only obvious
thing to
preserve is the zero velocity, so we consider all $g\in SU_2$ in the equation.

 However this is obviously impossible for non-zero $\psi$. Since $-g$ gives the
same
Lorentz transformation as $g$, we would need $g.\psi(p)=-g.\psi(p)$. In fact
much worse is true: There are many Lorentz transformations $g$ fixing $p$ and
the
$t$ axis. Then we need $g.\psi(p)=\psi(p)$ for such $g$.

However there is a way out, we consider the spinor field $\psi$ to be a
function on a circle bundle over space time. The circle will give our extra
spacelike coordinate $x^4$. It is very easy to
see what this bundle is: Over the unit sphere $S^2$ about the origin in spatial
coordinates we have $SU_2$ (which is
homeomorphic to $S^3$) and the map $g\mapsto g(0,1,0,0)$ is a Hopf fibering
$:SU_2\to S^2$. The full circle bundle
bundle is just the Hopf bundle times $\Bbb R^+\times \Bbb R$ (for the radius
and time variables), and is definitely
non-trivial over space time with the time axis removed. The bundle cannot be
extended to the time axis itself, so we
have a singularity there. Then the angular behaviour of $\psi$ is perfectly
well defined, and we only need to find the
radial and temporal behaviour. In addition, we will throw in a sixth timelike
dimension $x^5$, which will turn out to
generate mass. This will be useful in assigning quantum numbers to solutions.
To summarise the bundle structure, we
give the map from the extended $4+2$ dimensional space to the usual $3+1$
dimensional space-time:
$$\big(t,r,g,x^5\big) \quad\longmapsto\quad \big(t,g(r,0,0)\big)\quad g\in
SU_2,\ t,x^5\in \Bbb R,\ r\in (0,\infty)\
.$$

The reader may note that we have been somewhat liberal in our use of the
coordinate $x^4$. Since the Hopf bundle is
not topologically trivial, this coordinate cannot exist globally. However we
only need $\partial_4$, which has a well
defined meaning. If we fix a point in space, we can look towards the center of
the solution, and rotate  anticlockwise
about the axis defined by the two points. It is this rotation which generates
$\partial_4$ at that point. The size of the
rotation which gives $\partial_4$ is determined by the unique adjoint invariant
metric on $su_2$, to be the same as
the size of rotation giving a unit length vector in the physical 4-space. In
the same way, the derivatives
$\partial_2$ and $\partial_3$ can no longer be interpreted as partial
derivatives (as can be seen by the fact that they
do not commute). They are derivatives along orthogonal directions in the
tangent space to $SU_2$.

We assume that the potentials $A_{\mu}$ are actually well defined on the
original 4 dimensional space-time (so they
do not depend on our fiber coordinates $x^4$ and $x^5$).
 The hairy ball theorem says that the sphere has no global tangential vector
fields which do not vanish somewhere. In order not to break spherical symmetry,
we require that the angular components of the potentials vanish identically,
and that the temporal and radial components do not depend on angle.

\smallskip
To simplify the calculations, we will consider only the point $(0,r,0,0)$ in
$3+1$ space-time.
Having decided this, we find that $A^2$ and $A^3$ must vanish at our chosen
point $(0,r,0,0)$ by the last paragraph.
We must now reinterpret the meaning of $\partial_2$ and $\partial_3$. We take
$\partial_2$ to be the derivative along the Hopf bundle ($SU_2$) in a direction
which fixes the $z$ axis, and moves $(0,r,0,0)$ one unit in the $y$ direction.
This is implemented by the Lie algebra
element which has
all the $\omega^a_{\phantom{a}b}$
vanishing except
$\omega^2_{\phantom{2}1}=-\omega^1_{\phantom{1}2}=\frac 1r$.
Likewise we take
$\partial_3$ to be the derivative along the Hopf bundle in a direction
which fixes the $y$ axis, and moves $(0,r,0,0)$ one unit in the $z$ direction.
The resulting derivatives are
$$\partial_2 \psi\ =\ -\ \frac 1{4r}\big[ \gamma^2,\gamma^1\big] \psi \quad
\text{and}\quad
\partial_3 \psi\ =\ -\ \frac 1{4r}\big[ \gamma^3,\gamma^1\big]\psi\ .$$
The differential in the $x^4$ direction is given by the stabiliser of the point
$(0,r,0,0)$ in $SU_2$ as follows:
$$\partial_4 \psi\ =\  \frac 1{4r}\big[ \gamma^2,\gamma^3\big] \psi\ .$$
If we substitute this into the massless Dirac equation with
$A_2=A_3=A_4=A_5=0$, we find that
$$\gamma^0 \big(\partial_0-iqA_0\big)\psi\ +\
 \gamma^1 \big(\partial_1+\tfrac 1r -iqA_1\big) \psi\ +\ \frac
1{2r}\gamma^4\gamma^2\gamma^3\psi\ +\
\gamma^5\partial_5 \psi\ =\ 0\ .$$
 To simplify this we make the subsitiution $\phi=r\psi$, to find
$$\gamma^0 \big(\partial_0-iqA_0\big)\phi\ +\
 \gamma^1 \big(\partial_1 -iqA_1\big) \phi\ +\ \frac
1{2r}\gamma^4\gamma^2\gamma^3\phi\ +\
\gamma^5\partial_5 \phi\ =\ 0\ .$$
This can be rewritten as
$$\gamma^0 \big(-iqA_0+ \frac 1{2r}\gamma^0\gamma^4\gamma^2\gamma^3\big)\phi\
+\
 \gamma^1 \partial_1 \phi\ +\
\gamma^5\partial_5 \phi\ =\ 0\ ,$$
under the further assumptions that the solution is time independent, and that
$A_1=0$. The matrix
$\gamma^0\gamma^4\gamma^2\gamma^3$ is antihermitian, and we can take
eigenvectors $\phi_{\pm}$ with
eigenvalues $\gamma^0\gamma^4\gamma^2\gamma^3\phi_{\pm}=\pm i \phi_{\pm}$. Then
the following conditions
imply a solution to the Dirac equation:
$$A_0\ =\ \pm \frac{1}{2rq}\quad,\quad  A_{\mu}\ =\ 0 \quad(\mu\neq 0)
\quad,\quad  -\ \gamma^5\gamma^1\partial_1 \phi\ =\ \partial_5 \phi\ .$$
Note that the electromagnetic field is that of a point electric charge.
We take $x^5$ to parameterise a rescaling of the spinor field by $\partial_5
\phi=m\phi$ ($m>0$).
Then the Hermitian matrix $\gamma^5\gamma^1$ determines the sign of the mass. A
$+1$ eigenvector of the  matrix
$\gamma^5\gamma^1$ corresponds to a positive mass solution, while a
$-1$ eigenvector corresponds to a negative mass solution (i.e. exponentially
increasing with radial coordinate). Since
the matrices $\gamma^0\gamma^4\gamma^2\gamma^3$ and $\gamma^5\gamma^1$ commute,
we can choose
similtaneous eigenvectors.

\heading Section 3 \\The conserved currents.
\endheading

The conserved currents
$$j^{\mu}\ =\ q\psi^*\gamma^1\gamma^2\gamma^3\gamma^4\gamma^{\mu}\psi$$
are determined by the charge and mass of the solution. To have a definite
charge and mass, we take $\psi$ to be an
eigenvector of $\gamma^0\gamma^4\gamma^2\gamma^3$ and of $\gamma^5\gamma^1$. To
find the conserved
currents, we use the fact that $\psi^*\theta=0$ if $\psi$ and $\theta$ are from
different eigenspaces of a Hermitian
or anti-Hermitian matrix.

Now $\gamma^1\gamma^2\gamma^3\gamma^4\gamma^1$ and
$\gamma^1\gamma^2\gamma^3\gamma^4\gamma^5$ both anticommute with
$\gamma^0\gamma^4\gamma^2\gamma^3$, so they swap eigenspaces of
$\gamma^0\gamma^4\gamma^2\gamma^3$.
But these eigenspaces are perpendicular, so we find that $j^1=j^5=0$. For
example, if $\psi$ is a $+i$ eigenvector of
$\gamma^0\gamma^4\gamma^2\gamma^3$, then
$$(\gamma^0\gamma^4\gamma^2\gamma^3)(\gamma^1\gamma^2\gamma^3\gamma^4\gamma^{1})\psi\ =\ -
(\gamma^1\gamma^2\gamma^3\gamma^4\gamma^{1})(\gamma^0\gamma^4\gamma^2\gamma^3)\psi\ =\ -i
(\gamma^1\gamma^2\gamma^3\gamma^4\gamma^{1})\psi\ ,$$
so $\gamma^1\gamma^2\gamma^3\gamma^4\gamma^{1}\psi$ is from the $-i$ eigenspace
of
$\gamma^0\gamma^4\gamma^2\gamma^3$. This means that
$j^1=q\psi^*\gamma^1\gamma^2\gamma^3\gamma^4\gamma^{1}\psi=0$. Similarly
$\gamma^1\gamma^2\gamma^3\gamma^4\gamma^{\mu}$ anticommutes with
$\gamma^5\gamma^1$ for $\mu\neq
5,1$, so $j^{\mu}=0$ for $\mu\neq 5,1$. So the conserved currents vanish
identically (except at the
singularity itself). This explains why the electromagnetic field is that of a
point source, there is no charge density
outside the central point.

This gives rise to a scaling ambiguity in the solution, since Maxwell's
equation is identically satisfied for any scalar
multiple of the spinor field. This is effectively allowed for in the behaviour
of the spinor field as $x^5$ is increased.
S. Majid has mentioned that changing $x^5$ could be analogous to the action of
the renormalisation group.

The question of what happens at the central point should be addressed lest it
cause confusion. The solution is properly
currentless, as all the $j^{\mu}$ vanish for all space-time. The $x^0$ axis
(the world line of the electron) has to be
removed from space time, as there is no way to extend the Hopf bundle above it.
The electromagnetic field behaves as
though there were a point charge on the world line, but the currents cannot
possibly be defined there, as the spinor
field cannot be defined there. The Schwarzschild metric is considered as a
solution to the vacuum equations of general
relativity, despite the fact that it appears to an external observer that there
is a point mass at the center. People
correctly do not try to apply the equations at the singularity itself. But this
singularity is relatively mild compared to
the singularity in the Hopf fibering above the world line.

\heading Section 4 \\The quantum numbers.
\endheading

The matrix $\gamma^0\gamma^4\gamma^2\gamma^3$ which gives the sign of the
charge may look rather
asymmetric. However it is $\gamma^0\gamma^4(\text{stabiliser of the rotation
symmetry of the electron at the
point})$. As such, it takes a different form at different points in space-time
as the stabiliser of the point
changes. At a point on the $x^1$ axis it is just rotation about the $x^1$ axis.
In the same way
$\gamma^5\gamma^1$ has a description which is dependent on the point in
question.
 For simplicity, we continue to consider
a point on the $x^1$ axis rather than the general expression for an arbitrary
point.

The vectors $\psi$ take values in an 8 dimensional complex space, which is the
space of Dirac spinors in 6 space-time
dimensions. So far we have a 2 dimensional solution space for given mass and
charge (that is, for given eigenvalues of
$\gamma^0\gamma^4\gamma^2\gamma^3$ and $\gamma^5\gamma^1$). If we look for
another matrix commuting
with these quantities, we find that the stabilising rotation $\gamma^2\gamma^3$
will do. Then we can further
subdivide the solutions into spin states, that is $\pm i$ eigenspaces of the
stabiliser $\gamma^2\gamma^3$. It is
interesting that the spin states are distinguished by eigenstates of a
rotation.  We find
that any of the 8 possible combinations of $\pm$ mass,  $\pm$ spin,  and $\pm$
charge are allowed.  Here is the
formula for the solution:
$$\psi(t,r,g,x^5)\ =\ \frac 1r e^{\pm mr}e^{mx^5}.g\psi_0 \quad,\quad A_0\ =\
\pm \frac{1}{2rq}\ .$$
Here $g\in SU_2$ is a point in the Hopf bundle over the sphere, where the
identity in $SU_2$ is taken to lie over the
positive $x^1$ axis. The spinor component can be multiplied by any complex
scale factor. (A multiplication by a unit
norm number is a global gauge transformation, and multiplication by a positive
number is equivalent to shifting $x^5$).

The fact that the charge is inversely proportional to the coupling constant is
not really very suprising. The more
strongly interacting the theory, the easier it is to create singularities. The
existence of negative mass solutions may
seem worrying.  However, wheras the positive mass solutions are localised, the
spinor field for the negative mass
solutions increases indefnitely with radius. This makes it unlikely that such a
negative mass solution could be created
from interactions of positive mass particles on energetic grounds. Any sort of
reasonable boundary condition for the
field towards infinity would only allow positive mass particles.

\smallskip
The solutions for stationary electrons or positrons can be Lorentz transformed
to give moving solutions. A much more
complicated problem is how two such solutions would interact. An even bigger
question is whether the 6 dimensional
currentless equation is integrable [9]. If it were, it would provide a useful
stepping stone between the 2 dimensional
integrable models and 4 dimensional quantum field theories.

\heading References \endheading

(1) \quad M.F. Atiyah, `Geometry of Yang Mills fields',
Scuola Normale Superiore, Pisa (1979).

(2) \quad M.F. Atiyah, N.J. Hitchin, V.G. Drinfeld \& Yu.I. Manin,
`Construction of instantons',
Phys. Lett. A65, 185-187 (1978).

(3) \quad R. Coquereaux \& A. Jadczyk, `Geometry of multidimensional
universes',
Commun. Math. Phys. 90, 79-100 (1983).

(4) \quad P. Dirac, `Wave equations in conformal space', Ann. Math. 37 no. 2,
429-442  (1936).

(5) \quad P. Forg\'acs \& N.S. Manton, `Space-time symmetries in gauge
theories', Commun. Math. Phys. 72, 15-35 (1980).

(6) \quad P. Goddard \& D.I. Olive, `Magnetic monopoles in gauge field
theories', Rep. Prog. Phys. 41, 1357-1437  (1978).

(7) \quad S.V. Ketov, H. Nishino \& S.J. Gates, `Majorana-Weyl spinors and
self-dual gauge fields in 2+2 dimensions',
Physics Lett. B 307, 323-330 (1993).

(8) \quad E.C. Marino, `Are there topologically charged states associated with
Quantum Electrodynamics?', preprint
(1993).

(9) \quad R.S. Ward, `Completely solvable gauge-field theories in dimensions
greater than four', Nuclear Physics B 236,
 381-396 (1984).

(10) \quad C. Wetterich, `Chiral fermions in six-dimensional gravity', Nuclear
Physics B
253, 366-374 (1985).

(11) \quad C. Wetterich, `Fermion mass predictions from higher dimensions',
Nuclear Physics B
261, 461-490 (1985).

\enddocument